\documentstyle[aps]{revtex}

\begin{document}

\noindent \noindent \noindent \noindent \noindent \noindent {\bf RADIATION\
FROM\ PERFECT\ MIRRORS STARTING FROM REST AND ACCELERATING FOREVER AND THE
BLACK BODY SPECTRUM}

\bigskip \noindent {\it (revised version)}

\bigskip \noindent \noindent {\bf A\ Calogeracos}\noindent (*)

\noindent {\it Division of Theoretical Mechanics, Hellenic Air Force Academy
TG1010, Dhekelia Air Force Base, Greece}

\medskip

\noindent

\noindent

\noindent

\noindent \noindent 1st February 2002

\bigskip \noindent (*) acal@hol.gr

\medskip

\noindent

\begin{center}
\bigskip
\end{center}

{\bf Abstract}

\bigskip

We address the question of radiation emission from a perfect mirror that
starts from rest and follows the trajectory $z=-\ln (\cosh t)$ till $%
t\rightarrow \infty $. We show that a correct derivation of the black body
spectrum via the calculation of the Bogolubov amplitudes requires
consideration of the whole trajectory and not just of its asymptotic part.

\section{\protect\bigskip Introduction}

In a companion paper (Calogeracos 2002) hereafter referred to as I we
addressed the question of emission of radiation from accelerated mirrors
following prescribed relativistic asymptotically inertial trajectories. The
fact that the mirror moved at uniform velocity enabled us to define $in$ and 
$out$ states and employ standard time dependent perturbation theory. In the
present paper we consider a perfect mirror starting from rest and
accelerating for an infinite time along the trajectory 
\begin{equation}
z=g(t)=-\frac{1}{\kappa }\ln (\cosh \kappa t)  \label{bb1}
\end{equation}

\noindent The above problem has been considered extensively in the past,
starting with the classic papers by Fulling and Davies (1976) and Davies and
Fulling (1977) (DF\ in what follows). The problem is of interest because of
its (alleged) connection to radiation emitted from a collapsing black hole
(Hawking 1975), (DeWitt 1975) and to the attendant thermal spectrum. DF have
calculated the renormalized matrix element of the $T_{uu}$ component of the
energy momentum tensor, the latter being defined as 
\[
T_{uu}=\left( \partial _{u}\phi \right) ^{2}
\]
\noindent (see e.g. Birrell and Davies (1982), equation (4.16)). The result
is that {\it asymptotically for }$t\rightarrow \infty $%
\begin{equation}
\left\langle T_{uu}\right\rangle \rightarrow \frac{\kappa ^{2}}{48\pi }
\label{bb1a}
\end{equation}
\noindent To derive (\ref{bb1a}) DF have used the asymptotic expression for
the trajectory 
\begin{equation}
g(t)\rightarrow -t-Ae^{-2\kappa t}+B  \label{bb1b}
\end{equation}
\noindent where $A,B$ are constants that can be readily determined from (\ref
{bb1}). Equation (\ref{bb1a}) shows that there is a constant energy flux at
late times, which is interpreted (p. 249 of DF) as being analogous to the
thermal energy flux found by Hawking (1975) in the case of a black hole. DF
also calculated the Bogolubov amplitude $\beta (\omega ,\omega ^{\prime })$
(and thus $\alpha (\omega ,\omega ^{\prime })$ as well) and the spectrum 
\begin{equation}
n(\omega )=\int_{0}^{\infty }d\omega ^{\prime }\left| \beta (\omega ,\omega
^{\prime })\right| ^{2}e^{-a\left( \omega +\omega ^{\prime }\right) }
\label{bb2}
\end{equation}
\noindent ($a$ is a convergence factor). They did find that the spectrum
coincides with the black body spectrum, namely that 
\begin{equation}
\left| \beta (\omega ,\omega ^{\prime })\right| ^{2}=\frac{1}{2\pi \omega
^{\prime }}\frac{1}{e^{2\pi \omega }-1}  \label{bb1c}
\end{equation}
\noindent In this note we wish to re-examine the method used by DF to obtain
the expressions for the Bogolubov amplitudes. Our findings are summarized in
the concluding section, and there is no reason to anticipate them here. The
motivation for the investigation is provided by the following observations.
(a) The Bogolubov $\alpha (\omega ,\omega ^{\prime })$ and $\beta (\omega
,\omega ^{\prime })$ amplitudes are by definition time independent
quantities. Hence one should be sceptical as to the validity of using the
limiting form (\ref{bb1b}) ab initio without justification. (b) A careful
calculation reveals that a certain term is missing from the starting
equation (2.10a) of DF. Quite apart from the trivial nature of the error, it
is not clear that the omission does not undermine the validity of the final
result (\ref{bb1c}).

As far as the physics of the problem is concerned one may well counter that
in the context of mirrors acceleration continuing for an infinite time
implies mathematical singularities and also entails physical pathologies
associated, for example, with the infinite energy that has to be imparted to
the mirror. In that sense the realistic problem is examined in I. In the
present note we ignore such questions, take the premises of the early papers
on the subject for granted, and concentrate on the calculation of the
Bogolubov amplitudes and of the spectrum of the emitted radiation. \noindent
In what follows we will show that asymptotic behaviour of $\beta (\omega
,\omega ^{\prime })$ for large $\omega ^{\prime }$ is 
\begin{equation}
\beta (\omega ,\omega ^{\prime })\approx \left( \omega ^{\prime }\right) ^{-%
\frac{1}{2}}+O\left( \left( \omega ^{\prime }\right) ^{-N}\right) \left(
N>1\right)  \label{bb2a}
\end{equation}

\noindent The $\omega ^{\prime }$ integration in (\ref{bb2}) is then
logarithmically divergent. The divergence signifies according to DF (p. 250)
the production of particles at a finite rate for an infinite time. A similar
divergence has been noticed by Hawking (1975) (p. 211) in the context of
black holes and the interpretation offered is the same. One often refers to
the ultraviolet divergence by saying that large $\omega ^{\prime }$
frequencies dominate. In that same asymptotic limit $n(\omega )$ takes the
familiar form of the black body spectrum (times the logarithmically
divergent factor). Our analysis emphasizes two points: (a) the thermal
result depends crucially on the fact that asymptotically the trajectory
tends to a null line (the line $v=\ln 2$ in the case of (\ref{bb1})), (b)
for a mirror starting from rest and forever accelerating one must consider
the {\it whole }trajectory and {\it not }just the asymptotic portion of it.
The truth of statement (a) is usually taken as common knowledge. However
statement (a) is also taken erroneously to imply the negative of (b).

Note: Notation and conventions follow that of I (and are largely in
accordance with the papers cited in the Introduction). In section 2 we
briefly go through the standard notation so that the paper be self contained.

\section{The formalism for a perfect mirror accelerating forever}

We introduce the usual light cone coordinates 
\begin{equation}
u=t-z,v=t+z  \label{e01}
\end{equation}

\noindent The massless Klein-Gordon equation reads 
\begin{equation}
{\displaystyle {\partial ^{2}\phi  \over \partial u\partial v}}%
=0  \label{e1}
\end{equation}

\noindent Hence any function that depends only on $u$ or $v$ (or the sum of
two such functions) is a solution of (\ref{e1}). Let 
\begin{equation}
z=g(t)=-\ln (\cosh t)  \label{bb3}
\end{equation}
be the mirror's trajectory. Notice that for large $t$ the trajectory assumes
the asymptotic form 
\[
g(t)\approx -t-e^{-2t}+\ln 2 
\]
Details are given in Appendix A of I; for the moment observe that the
trajectory is asymptotic to the line $v=\ln 2$ (figure 7 of I). In terms of
the $u,$ $v$ coordinates the equation (\ref{bb3}) of the trajectory is
written in the form 
\begin{equation}
u=f(v)  \label{e005}
\end{equation}
or alternatively 
\begin{equation}
v=p(u)  \label{e05}
\end{equation}
where the function $f$ is the inverse of $p$. The construction of the
functions $f$ and $p$ given the equation of the trajectory $z=g(t)$ is
explained in section 2 of I. In the particular case of (\ref{bb3}) the
function $f$ is given by 
\[
f(v)=1,v\leq 0 
\]
\begin{equation}
f(v)=-\ln \left( 2-e^{v}\right) ,0\leq v<\ln 2  \label{208b}
\end{equation}

\noindent For future use we quote 
\begin{equation}
f^{\prime }(v)=-\frac{e^{v}}{2-e^{v}}  \label{bb11}
\end{equation}

We take everything to exist to the right of the mirror. One set of modes
satisfying (\ref{e1}) and the boundary condition 
\begin{equation}
\phi (t,g(t))=0  \label{e2}
\end{equation}

\noindent is given by (see I, equation (4)) 
\begin{equation}
\varphi _{\omega }(u,v)=%
{\displaystyle {i \over 2\sqrt{\pi \omega }}}%
\left( \exp (-i\omega v)-\exp \left( -i\omega p(u)\right) \right)  \label{e3}
\end{equation}

\noindent Another set of modes satisfying the boundary condition is
immediately obtained from (\ref{e3}) 
\begin{equation}
\bar{\varphi}_{\omega }(u,v)=%
{\displaystyle {i \over 2\sqrt{\pi \omega }}}%
\left( \exp \left( -i\omega f(v)\right) -\exp \left( -i\omega u\right)
\right)  \label{e5}
\end{equation}

\noindent

The modes $\varphi _{\omega }(u,v)$ of (\ref{e3}) describe waves incident
from the right as it is clear from the sign of the exponential in the first
term; the second term represents the reflected part which has a rather
complicated behaviour depending on the motion of the mirror. These modes
constitute the {\it in }space and should obviously be absent before
acceleration starts, 
\begin{equation}
a(\omega ^{\prime })\left| 0in\right\rangle =0  \label{bb4}
\end{equation}
Similarly the modes $\bar{\varphi}_{\omega }(u,v)$ describe waves travelling
to the right (emitted by the mirror) as can be seen from the exponential of
the second term. Correspondingly the first term is complicated. Regarding
the {\it out }modes (\ref{e5}) the following point, which will be useful in
the evaluation of the matrix element $\beta (\omega ,\omega ^{\prime })$ (%
\ref{e11}), ought to be noted. Recall the picturesque form of such a mode
(figure 6 of I), where for a perfect mirror $\bar{R}_{R}=1,$ $\bar{T}_{R}=0$%
. Clearly in the coordinate range $u\leq 0,v\leq 0$ the mirror is stationary
and thus the modes trivialize. During the subsequent motion of the mirror
the variable $v$ in the back-scattered wave in (\ref{e5}) is restricted
between $0$ and $\ln 2$. These remarks will reflect on the range of
integration in (\ref{e100}).

The $\bar{\varphi}_{\omega }(u,v)$ modes define the {\it out }space and 
\begin{equation}
\bar{a}(\omega )\left| 0out\right\rangle =0  \label{bb6}
\end{equation}
The state \noindent $\left| 0out\right\rangle $ corresponds to the state
where nothing is produced by the mirror. The two representations are
connected by the Bogolubov transformation 
\begin{equation}
\bar{a}\left( \omega \right) =\int_{0}^{\infty }d\omega \left( \alpha
(\omega ,\omega ^{\prime })a(\omega ^{\prime })+\beta ^{*}(\omega ,\omega
^{\prime })a^{\dagger }(\omega ^{\prime })\right)  \label{e011}
\end{equation}

\noindent Using (\ref{e011}) and its hermitean conjugate we may immediately
verify that the expectation value of the number of excitations of the mode $%
\left( \omega \right) $ in the $\left| 0in\right\rangle $ vacuum is given by 
\begin{equation}
\left\langle 0in\right| N\left( \omega \right) \left| 0in\right\rangle
=\int_{0}^{\infty }d\omega ^{\prime }\left| \beta (\omega ,\omega ^{\prime
})\right| ^{2}  \label{e0011}
\end{equation}

\noindent The matrix element $\beta (\omega ,\omega ^{\prime })$ is given by
(Birrell and \ Davies 1982), equations (2.9), (3.36) 
\begin{equation}
\beta (\omega ,\omega ^{\prime })=-i\int dz\varphi _{\omega ^{\prime }}(z,0)%
{\displaystyle {\partial  \over \partial t}}%
\bar{\varphi}_{\omega }(z,0)+i\int dz\left( 
{\displaystyle {\partial  \over \partial t}}%
\varphi _{\omega ^{\prime }}(z,0)\right) \bar{\varphi}_{\omega }(z,0)
\label{e11}
\end{equation}

\noindent The integration in (\ref{e11}) can be over any spacelike
hypersurface. Since the mirror is at rest for $t\leq 0$ the choice $t=0$ for
the hypersurface is convenient. The $in$ modes evaluated at $t=0$ are given
by the simple expression 
\begin{equation}
\varphi _{\omega }(u,v)=%
{\displaystyle {i \over 2\sqrt{\pi \omega }}}%
\left( \exp (-i\omega v)-\exp \left( -i\omega u\right) \right)  \label{e99}
\end{equation}
i.e. expression (\ref{e3}) with $p(u)=u$ (corresponding to zero velocity).
The $\bar{\varphi}$ modes are given by (\ref{e5}) with $f$ depending on the
trajectory. Relation (\ref{e11}) is rewritten in the form (the endpoints of
integration shall be stated presently) 
\begin{eqnarray}
\beta (\omega ,\omega ^{\prime }) &=&-i\int dz%
{\displaystyle {i \over 2\sqrt{\pi \omega ^{\prime }}}}%
\left\{ e^{-i\omega ^{\prime }z}-e^{i\omega ^{\prime }z}\right\} \frac{%
\omega }{2\sqrt{\pi \omega }}\left\{ f^{\prime }e^{-i\omega f}-e^{i\omega
z}\right\} +  \label{e100} \\
&&+i\int dz%
{\displaystyle {\omega ^{\prime } \over 2\sqrt{\pi \omega ^{\prime }}}}%
\left\{ e^{-i\omega ^{\prime }z}-e^{i\omega ^{\prime }z}\right\} \frac{i}{2%
\sqrt{\pi \omega }}\left\{ e^{-i\omega f}-e^{i\omega z}\right\}  \nonumber
\end{eqnarray}

\noindent Bearing in mind the comments preceding (\ref{bb6}) regarding the
range of integration, the above expression may be rearranged in the form 
\begin{equation}
\beta (\omega ,\omega ^{\prime })=%
{\displaystyle {1 \over 4\pi \sqrt{\omega \omega ^{\prime }}}}%
\int_{0}^{\ln 2}dz\left\{ e^{i\omega ^{\prime }z}-e^{-i\omega ^{\prime
}z}\right\} \left\{ \omega ^{\prime }e^{-i\omega f}-\omega f^{\prime
}e^{-i\omega f}\right\} + 
{\displaystyle {\left( \omega -\omega ^{\prime }\right)  \over 4\pi \sqrt{\omega \omega ^{\prime }}}}%
\int_{0}^{\infty }dz\left\{ e^{i\omega ^{\prime }z}-e^{-i\omega ^{\prime
}z}\right\} e^{i\omega z}  \label{bb9}
\end{equation}

\noindent where the limits of integration are displayed. The first and
second integral in the above relation will be denoted by $\beta _{I}(\omega
,\omega ^{\prime })$ and $\beta _{III}(\omega ,\omega ^{\prime })$
respectively (in accordance with the notation in I). Notice that the second
integration is of kinematic origin and independent of the form of the
trajectory. We also quote the expression for the $\alpha (\omega ,\omega
^{\prime })$ amplitude 
\begin{equation}
\alpha (\omega ,\omega ^{\prime })=i\int_{0}^{\infty }dz\varphi _{\omega
^{\prime }}(z,0)%
{\displaystyle {\partial  \over \partial t}}%
\bar{\varphi}_{\omega }^{*}(z,0)-i\int_{0}^{\infty }dz\left( 
{\displaystyle {\partial  \over \partial t}}%
\varphi _{\omega ^{\prime }}(z,0)\right) \bar{\varphi}_{\omega }^{*}(z,0)
\label{91}
\end{equation}

\noindent Observe the unitarity condition (see I for details) 
\begin{equation}
\int_{0}^{\infty }d\widetilde{\omega }\left( \alpha \left( \omega _{1},%
\widetilde{\omega }\right) \alpha ^{*}\left( \omega _{2},\widetilde{\omega }%
\right) -\beta \left( \omega _{1},\widetilde{\omega }\right) \beta
^{*}\left( \omega _{2},\widetilde{\omega }\right) \right) =\delta \left(
\omega _{1}-\omega _{2}\right)  \label{92}
\end{equation}

\noindent which is a consequence of the fact that the set of {\it in }states
is complete. Notice that identity (69) of I is not valid because the set of 
{\it out} states is not complete (in the present case where the trajectory
has a $v$ asymptote). On the question of completeness see the remarks
preceding equation (35) of I.

Recall from I that we can introduce quantities $A(\omega ,\omega ^{\prime
}),B(\omega ,\omega ^{\prime })$ that are analytic functions of the
frequencies via 
\begin{equation}
\alpha (\omega ,\omega ^{\prime })=\frac{A(\omega ,\omega ^{\prime })}{\sqrt{%
\omega \omega ^{\prime }}},\beta (\omega ,\omega ^{\prime })=\frac{B(\omega
,\omega ^{\prime })}{\sqrt{\omega \omega ^{\prime }}}  \label{e106}
\end{equation}

\noindent The quantity $B(\omega ,\omega ^{\prime })$ is read off (\ref{e100}%
) (and $A(\omega ,\omega ^{\prime })$ from the corresponding expression for $%
\alpha (\omega ,\omega ^{\prime })$). From the definitions of the Bogolubov
coefficients, the explicit form (\ref{e100}) of the overlap integral and
expressions (\ref{e3}) and (\ref{e5}) for the field modes one can deduce
that 
\begin{equation}
B^{*}(\omega ,\omega ^{\prime })=A(-\omega ,\omega ^{\prime }),A^{*}(\omega
,\omega ^{\prime })=B(-\omega ,\omega ^{\prime })  \label{e107}
\end{equation}

\noindent The above relations allow the calculation of $\alpha (\omega
,\omega ^{\prime })$ once $\beta (\omega ,\omega ^{\prime })$ is determined.

\section{Comparison between realistic trajectories and trajectories
accelerating forever}

A mirror trajectory involving acceleration forever cannot be considered as
realistic in contrast to an asymptotically inertial trajectory. The former
type of trajectory is of interest in that it can be taken as a simplified
analog of black hole collapse. The two trajectories are radically different
from a physical point of view and this is reflected in the form of their
spectra. In I we considered a mirror starting from rest, accelerating along
the trajectory (\ref{bb1}) till a spacetime point {\it P} (labelled by a
coordinate $v=r<\ln 2$ in $(u,v)$ coordinates), and then continuing at
uniform velocity. We showed that in that case the Bogolubov amplitude
squared $\left| \beta (\omega ,\omega ^{\prime })\right| ^{2}$ behaves
asymptotically for large $\omega ^{\prime }$ as $\left( \omega ^{\prime
}\right) ^{-5}$. In this work we set out to show that if acceleration is to
continue forever then $\left| \beta (\omega ,\omega ^{\prime })\right| ^{2}$
goes as $1/\omega ^{\prime }$ (see (\ref{bb2a})). It is thus not the case
that the second problem can be simply considered as the $r\rightarrow \ln 2$
limit of the first. Although this may not come as a surprise from a physical
point of view, it is certainly interesting to see where the dichotomy occurs
mathematically. To this end we return to the problem examined in I where the 
$\beta (\omega ,\omega ^{\prime })$ amplitude is given by (\ref{bb9}) above
with $\ln 2$ replaced by $\infty $ (cf also equation (34) of I), with the
understanding that the function $f(z)$ stands for $f_{acc}(z)$ given by (\ref
{208b}) above when $z<r,$ and for $f_{0}(z)$ given by (13) of I when $z>r$
(in the last equation $B$ stands for the velocity $\beta _{P}$ at point {\it %
P} where acceleration stops and the constant $C$ is adjusted so that
velocity be continuous at {\it P}; see equation (49) of I). We introduce the
Fourier integrals 
\begin{equation}
I_{1}=\int_{0}^{\infty }dze^{i\omega ^{\prime }z-\alpha z}e^{-i\omega f(z)}
\label{t1}
\end{equation}

\begin{equation}
I_{2}=\int_{0}^{\infty }dze^{i\omega ^{\prime }z-\alpha z}f^{\prime
}(z)e^{-i\omega f(z)}  \label{t2}
\end{equation}

\[
I_{3}=\int_{0}^{\infty }dze^{-i\omega ^{\prime }z-\alpha z}e^{-i\omega f(z)} 
\]

\[
I_{4}=\int_{0}^{\infty }dze^{-i\omega ^{\prime }z-\alpha z}f^{\prime
}(z)e^{-i\omega f(z)} 
\]

\noindent where $\alpha $ is a small convergence factor taken to zero at the
end of the calculations. Thus the $f$-dependent part of amplitude (35) of I
is a linear combination of the above four integrals. The treatment that
follows is equivalent to that of section 3.A of I, however it makes clearer
the contrast between the two cases (realistic trajectory vs one that
accelerates forever).

To examine the asymptotic behaviour for large $\omega ^{\prime }$ of the
above integrals we use simple integration by parts (see Bender and Orszag
1978), p. 278 according to the formula 
\begin{equation}
\int_{a}^{b}dzF(z)e^{i\omega ^{\prime }z}=\left[ \frac{F(z)}{i\omega
^{\prime }}e^{i\omega ^{\prime }z}\right] _{a}^{b}-\frac{1}{i\omega ^{\prime
}}\int_{a}^{b}dzF^{\prime }(z)e^{i\omega ^{\prime }z}  \label{t3}
\end{equation}

\noindent which works provided the quantities appearing in the right hand
side are well defined. To treat $I_{1}$ of (\ref{t1}) we split the integral $%
\int_{0}^{\infty }$ to $\int_{0}^{r}+\int_{r}^{\infty }$ and apply (\ref{t3}%
) twice. The endpoint contributions from infinity vanish due to the
convergence factor. The endpoint contributions at $z=r$ cancel out in pairs: 
$f_{0}(r)$ cancels with $f_{acc}(r)$ (both accompanied by a factor $1/\omega
^{\prime }$) and $f_{0}^{\prime }(r)$ cancels with $f_{acc}^{\prime }(r)$
(both accompanied by a factor $\left( \omega ^{\prime }\right) ^{-2}$). The
origin of these cancellations is a direct consequence of the nature of the
trajectory considered. The first cancellation reflects the fact that the
trajectory itself $u=f(v)$ is continuous, and the second reflects the
continuity of the velocity at {\it P} (see equation (15) of I for the
general connection between $f^{\prime }(v)$ and velocity). Similarly
endpoint contributions at $z=0$ cancel with corresponding terms originating
from a large $\omega ^{\prime }$ expansion of the second (trajectory
independent) term in (\ref{bb9}) on the same continuity grounds as above.
Thus integral $I_{1}$ goes as $\left( \omega ^{\prime }\right) ^{-3}$ and
observing the prefactors in (\ref{bb9}) we deduce that its contribution to $%
\beta (\omega ,\omega ^{\prime })$ goes as $\left( \omega ^{\prime }\right)
^{-5/2}$.

The integral $I_{2}$ given by (\ref{t2}) can be handled in a similar way.
However because of the presence of $f^{\prime }(z)$ we can integrate by
parts only once if the endpoint contributions are to cancel out (recall that 
$f_{0}^{\prime \prime }(r)=f_{acc}^{\prime \prime }(r)$ would require
continuity of the acceleration). Thus $I_{2}$ goes as $\left( \omega
^{\prime }\right) ^{-2}$. Because of the corresponding prefactor in (\ref
{bb9}) is one power of $\omega ^{\prime }$ smaller, the final contribution
to the amplitude is again of the order $\left( \omega ^{\prime }\right)
^{-5/2}$.

We return to (\ref{bb9}) with the upper limit of the first integral now
being equal to $\ln 2$. It is manifestly obvious that there is now no room
for the subtle cancellations exhibited above. We can still try integration
by parts to see whether we can possibly arrive at any conclusions. To this
end consider (for example) $I_{2}$ (with the upper limit now equal to $\ln
2).$ Recall that according to (\ref{208b}) 
\[
e^{-i\omega f(z)}=e^{-i\omega \ln (2-e^{z})}=\left( 2-e^{z}\right)
^{-i\omega } 
\]

\noindent and according to (\ref{bb11}) 
\[
f^{\prime }(z)=-\frac{e^{z}}{2-e^{z}} 
\]

\noindent Then 
\[
I_{2}=-\int_{0}^{\ln 2}dze^{-i\omega ^{\prime }z}\frac{e^{z}}{\left(
2-e^{z}\right) ^{1+i\omega }} 
\]

\noindent If we now attempt to apply property (\ref{t3}) the upper endpoint
contribution diverges. Hence we are unable to extract a further power of $%
1/\omega ^{\prime }$. It is clear that to obtain an asymptotic estimate in
the present problem we have to resort to other methods.

\section{Calculation of the Bogolubov amplitudes}

The strategy we adopt in handling (\ref{bb9}) is as follows. The first
integral will be evaluated via an asymptotic expansion in negative powers of 
$\omega ^{\prime }$, which will in fact show that the $\omega ^{\prime }$
integration in (\ref{bb2}) is logarithmically divergent. The second integral 
$\beta _{III}(\omega ,\omega ^{\prime })$ in (\ref{bb9}) is readily
evaluated (see also I) 
\begin{equation}
\beta _{III}(\omega ,\omega ^{\prime })=\frac{1}{4\pi i\sqrt{\omega \omega
^{\prime }}}-\frac{1}{4\pi i\sqrt{\omega \omega ^{\prime }}}\left( \omega
-\omega ^{\prime }\right) \zeta \left( \omega +\omega ^{\prime }\right)
\label{e103}
\end{equation}

\noindent where the function $\zeta $ and its complex conjugate $\zeta ^{*}$%
are defined in Heitler (1954), pages 66-71: 
\begin{equation}
\zeta (x)\equiv -i\int_{0}^{\infty }e^{i\kappa x}d\kappa =P\frac{1}{x}-i\pi
\delta (x)  \label{delta}
\end{equation}

\noindent As explained in I it is only the first term in (\ref{delta}) that
is operative as far as the calculation of the $\beta (\omega ,\omega
^{\prime })$ goes. (The $\delta $ proportional term is only relevant in the
calculation of the $\alpha (\omega ,\omega ^{\prime })$ amplitude via
relations (\ref{e106}), (\ref{e107}).) Thus asymptotically in the large $%
\omega ^{\prime }$ limit 
\begin{equation}
\beta _{III}(\omega ,\omega ^{\prime })\approx \frac{1}{2\pi i\sqrt{\omega
\omega ^{\prime }}}  \label{bb10}
\end{equation}

We turn to the first integral $\beta _{I}(\omega ,\omega ^{\prime })$ in (%
\ref{bb9}). Rather than dealing with four integrals we perform an
integration by parts to get (\ref{bb9}) in the form 
\begin{equation}
\beta _{I}(\omega ,\omega ^{\prime })=-\frac{1}{2\pi }\sqrt{\frac{\omega
^{\prime }}{\omega }}\int_{0}^{\ln 2}dze^{-i\omega f(z)-i\omega ^{\prime }z}+%
\frac{1}{2\pi }\frac{1}{\sqrt{\omega \omega ^{\prime }}}\sin \left( \omega
\ln 2\right) e^{-i\omega f\left( \ln 2\right) }  \label{bb22}
\end{equation}
This is the same integration by parts that was used to obtain (54) of I and
also the one that is used in DF to go from their (2.10a) to (2.10b). The
second term in (\ref{bb22}) oscillates rapidly since the exponent tends to
infinity. Thus the term tends distributionally to zero and it may be
neglected as in DF (also it is one power of $\omega ^{\prime }$ down
compared to the first term). Of course this term was kept in (47) of I since
it gives a finite contribution for $r<\ln 2$ (recall that at $v=r$ the
trajectory considered in I reverts to uniform velocity). So asymptotically
we are entitled to write 
\begin{equation}
\beta _{I}(\omega ,\omega ^{\prime })\approx -\frac{1}{2\pi }\sqrt{\frac{%
\omega ^{\prime }}{\omega }}\int_{0}^{\ln 2}dze^{-i\omega f(z)-i\omega
^{\prime }z}=-\frac{1}{2\pi }\sqrt{\frac{\omega ^{\prime }}{\omega }}%
\int_{0}^{\ln 2}dze^{-i\omega ^{\prime }z}\left( 2-e^{z}\right) ^{i\omega }
\label{bb23}
\end{equation}
To bring the singularity to zero we make the change of variable

\begin{equation}
z=\ln 2-\rho  \label{e22a}
\end{equation}

\noindent and rewrite $\beta _{I}(\omega ,\omega ^{\prime })$ in the form 
\begin{equation}
\beta _{I}(\omega ,\omega ^{\prime })\approx -\frac{2^{i\left( \omega
-\omega ^{\prime }\right) }}{2\pi }\sqrt{\frac{\omega ^{\prime }}{\omega }}%
\int_{0}^{\ln 2}d\rho e^{i\omega ^{\prime }\rho }\left( 1-e^{-\rho }\right)
^{i\omega }  \label{bb24}
\end{equation}

\noindent \noindent We isolate the integral

\begin{equation}
I\equiv \int_{0}^{\ln 2}d\rho e^{i\omega ^{\prime }\rho }\left( 1-e^{-\rho
}\right) ^{i\omega }  \label{e24b}
\end{equation}

\noindent

To obtain the asymptotic behaviour of (\ref{e24b}) we adopt the standard
technique of deforming the integration path to a contour in the complex
plane (see (Bender and Orszag 1978), chapter 6); see also (Morse and
Feshbach 1953), p. 610 where a very similar contour is used in the study of
the asymptotic expansion of the confluent hypergeometric. The deformed
contour runs from 0 up the imaginary axis till $iT$ (we eventually take $%
T\rightarrow \infty $), then parallel to the real axis from $iT$ to $iT+\ln
2 $, and then down again parallel to the imaginary axis from $iT+\ln 2$ to $%
\ln 2$. The contribution of the segment parallel to the real axis vanishes
exponentially in the limit $T\rightarrow \infty $. We thus get 
\[
I=i\int_{0}^{\infty }dse^{-\omega ^{\prime }s}\left( 1-e^{-is}\right)
^{i\omega }-i\int_{0}^{\infty }dse^{i\omega ^{\prime }\left( \ln 2+is\right)
}\left( 1-e^{-\ln 2-is}\right) ^{i\omega }= 
\]
\begin{equation}
=i\int_{0}^{\infty }dse^{-\omega ^{\prime }s}\left( 1-e^{-is}\right)
^{i\omega }-i2^{i\omega ^{\prime }}\int_{0}^{\infty }dse^{-\omega ^{\prime
}s}\left( 1-\frac{e^{-is}}{2}\right) ^{i\omega }  \label{bb13}
\end{equation}
In the limit of $\omega ^{\prime }$ large the main contribution to (\ref
{bb13}) comes from the $s\approx 0$ region. We expand $e^{is}$ in powers of $%
s$. We thus approximate one factor of the first integrand in (\ref{bb13}) by 
\[
1-e^{-is}\approx is,\left( is\right) ^{i\omega }=e^{-\pi \omega
/2}s^{i\omega } 
\]

\noindent where we set 
\begin{equation}
is=\left| s\right| e^{i\pi /2}  \label{bb13a}
\end{equation}

\noindent ,took the branch cut of the function $\rho ^{i\omega }$ to run
from zero along the negative $x$ axis, wrote $\rho ^{i\omega }=\exp \left(
i\omega \left( \ln \rho +i2N\pi \right) \right) $ and chose the branch $N=0$%
. For the second integrand we get 
\[
\left( 1-\frac{e^{-is}}{2}\right) ^{i\omega }\approx \left( \frac{1}{2}+%
\frac{is}{2}\right) ^{i\omega } 
\]

\noindent Thus

\begin{eqnarray}
I &=&ie^{-\pi \omega /2}\int_{0}^{\infty }dse^{-\omega ^{\prime
}s}s^{i\omega }-i2^{i\left( \omega ^{\prime }-\omega \right)
}\int_{0}^{\infty }dse^{-\omega ^{\prime }s}=  \label{bb14} \\
&=&ie^{-\pi \omega /2}\frac{\Gamma \left( 1+i\omega \right) }{\left( \omega
^{\prime }\right) ^{1+i\omega }}-\frac{i2^{i\left( \omega ^{\prime }-\omega
\right) }}{\omega ^{\prime }}  \nonumber
\end{eqnarray}

\noindent We substitute (\ref{bb14}) in (\ref{bb24}) and get 
\begin{equation}
\beta _{I}(\omega ,\omega ^{\prime })\approx -i\frac{2^{i\left( \omega
-\omega ^{\prime }\right) }}{2\pi \sqrt{\omega \omega ^{\prime }}}\left(
\omega ^{\prime }\right) ^{-i\omega }e^{-\pi \omega /2}\Gamma \left(
1+i\omega \right) +\frac{i}{2\pi \sqrt{\omega \omega ^{\prime }}}
\label{bb16}
\end{equation}

\noindent Observe the crucial cancellation of the second term in (\ref{bb16}%
) with (\ref{bb10}), the end result being 
\begin{equation}
\beta (\omega ,\omega ^{\prime })\approx -i\frac{2^{i\left( \omega -\omega
^{\prime }\right) }}{2\pi \sqrt{\omega \omega ^{\prime }}}\left( \omega
^{\prime }\right) ^{-i\omega }e^{-\pi \omega /2}\Gamma \left( 1+i\omega
\right)  \label{bb17}
\end{equation}

\noindent As already mentioned at the end of the previous section, having
determined $\beta (\omega ,\omega ^{\prime })$ allows one to determine $%
\alpha (\omega ,\omega ^{\prime })$. This is the typical form of the $\beta
(\omega ,\omega ^{\prime })$ amplitude that leads to the thermal spectrum;
see e.g. DF. Indeed by taking the modulus of (\ref{bb17}), squaring and
using the property 
\[
\left| \Gamma \left( 1+iy\right) \right| ^{2}=\pi y/\sinh \left( \pi
y\right) 
\]

\noindent we get the black body spectrum (\ref{bb1c}).

\section{Conclusion}

The objective of the paper was to prove that the Bogolubov amplitude $\beta
(\omega ,\omega ^{\prime })$ has the asymptotic form (\ref{bb2a}) and that
the radiation emitted has the spectrum of a black body. Before enlarging on
the conclusions we should elucidate one technical point in connection to the
classic paper by Davies and \ Fulling (1977). It seems that the term $\beta
_{III}(\omega ,\omega ^{\prime })$ is unaccountably missing from (2.10a) of
DF. Its existence is necessary if the unitarity conditions are to be
satisfied. The need for the $\beta _{III}$ term may be seen in a trivial
example based on the results of I. Consider the limiting case of a mirror
perpetually at rest. Then $\beta _{I}(\omega ,\omega ^{\prime })$ ((52) of
I) disappears (formally $r=0$) and $\beta _{III}$ is instrumental in
cancelling $\beta _{II}$ of (53) evaluated at $r=0$, $\beta _{P}=0$ so that
the total emission amplitude $\beta $ vanishes (as it has to). The omission
has been pointed out by Walker (1985) who however did not pursue the matter
any further. On the other hand the present derivation shows that its
existence is {\it crucial} in cancelling the non-thermal part of the $\beta
_{I}(\omega ,\omega ^{\prime })$ term.

There are various arguments in the literature in favour of the black body
spectrum in the case of trajectory (\ref{bb1}). In the present note we are
concentrating on a proof based on the Bogolubov coefficients. These
quantities are by definition time-independent, and in this context the
question as to where and when the photons are produced simply does not
arise. In the same vein it is totally arbitrary to assert from the start
that one specific part of the trajectory (in the present case the one near
the asymptote) is more important than other parts. It is certainly true that
were it not for the singularity on the $v$ asymptote the thermal spectrum
would not arise. However the main conclusion of this note is that the
correct derivation of the thermal result requires the consideration of the
complete trajectory and not just of its asymptotic part. In technical terms
the function $f(z)$ in (\ref{bb23}) cannot be approximated by its asymptotic
expression. These remarks are strengthened by (i) the aforementioned role of
the missing term (which appears {\it as if } it originates at $t=0$), (ii)
the failure of the short distance expansion from the $v$ asymptote as
demonstrated in the Appendix. All this is in accordance with one's quantum
mechanical intuition. One's classical instincts might dictate that roughly
speaking the small amount of time spent near the origin would have an
insignificant effect compared to the infinitely long time spent near the
asymptote. However such loose statements are misleading in connection to the
quantum mechanical calculation of global ({\it time independent) }%
quantities. Similarly attempts to distinguish between ''transient'' and
''steady state'' radiation at the level of the $\alpha $ and $\beta $
amplitudes are bound to fail; the emphasis in the literature on the
importance of the asymptotic part of the trajectory has unfortunately led to
such statements. Having said all that, one can certainly enquire about the
matrix elements of {\it local }field quantities as it is indeed done in DF
in a most illuminating way. This is however quite distinct from the
calculation of Bogolubov amplitudes.

Mention must be made to the work of Carlitz and Willey (1987) where a mirror
accelerating from the infinite past to the infinite future is considered.
The authors do get the black body spectrum, their amplitudes do satisfy the
Bogolubov identities, and quite clearly the time $t=0$ plays no special role
in their problem which is quite different from ours and our comments do not
apply to their work.

\noindent \noindent {\bf Acknowledgments}

The author is indebted to Professor S\ A\ Fulling, Dr G\ Plunien, Dr R\
Schutzhold, and Professor G\ E Volovik for discussions and/or \noindent
correspondence, and to two referees for their comments and suggestions. He
also wishes to thank Professor M\ Paalanen and the Low Temperature
Laboratory, Helsinki University of Technology, and Professor G\ Soff and the
Institute for Theoretical Physics, Dresden Technical University for their
hospitality.

\noindent {\bf Appendix: Short distance expansion from the horizon and large
frequencies}

Let us consider the integral (\ref{e24b}) and expand in small $\rho $. To
first order 
\begin{equation}
I^{(1)}=\int_{0}^{\ln 2}d\rho \rho ^{i\omega }e^{i\omega ^{\prime }\rho }
\label{680a}
\end{equation}

\noindent The integral (\ref{680a}) can be performed exactly in terms of the
confluent hypergeometric and the asymptotic limit of large $\omega ^{\prime
} $ may be examined afterwards. Let us make the change of variable $\rho
=t\ln 2$ in (\ref{680a}) and rewrite 
\begin{eqnarray}
I^{(1)} &=&\left( \ln 2\right) ^{i\omega +1}\int_{0}^{1}dte^{i\omega
^{\prime }t\ln 2}t^{i\omega }=  \label{bb18} \\
&=&\left( \ln 2\right) ^{i\omega +1}\frac{1}{i\omega +1}M\left( 1+i\omega
,2+i\omega ,i\omega ^{\prime }\ln 2\right)  \nonumber
\end{eqnarray}
(where $M$ is the confluent hypergeometric function). An alternative way of
presenting the above result is in terms of the incomplete gamma function $%
\gamma ^{*}$ by exploiting the connection of the latter with the confluent
(see e.g. Tricomi (1954)) 
\begin{equation}
\gamma ^{*}(\alpha ,z)=\frac{M(\alpha ,\alpha +1;-z)}{\Gamma (\alpha +1)}
\label{c3}
\end{equation}
We can now examine the asymptotic limit of (\ref{bb18}) for large $\omega
^{\prime }$. The asymptotic limit of the confluent $M(a,b,i\left| z\right| )$
for large values of $\left| z\right| $ is given by item 13.5.1 of
(Abramowitz and Stegun 1972) ($z\equiv i\omega ^{\prime }\ln 2$). In the
case $b=a+1$ some simplifications occur and we get 
\begin{equation}
M\left( 1+i\omega ,2+i\omega ,i\left| z\right| \right) \approx -\left(
1+i\omega \right) e^{i\left| z\right| }\frac{i}{\left| z\right| }+i\Gamma
\left( 2+i\omega \right) \frac{e^{-\frac{\pi \omega }{2}}}{\left| z\right|
^{1+i\omega }}  \label{bb20}
\end{equation}
(other terms are down by higher powers of $1/\left| z\right| $). The second
term of the above relation combined with the prefactors of (\ref{bb18}) does
feature the $\Gamma \left( 1+i\omega \right) e^{-\frac{\pi \omega }{2}}$
factor characteristic of the black body spectrum. However the presence of
the first term spoils the thermal result (notice that both terms are of the
same order in $\omega ^{\prime }$). In other words the result (\ref{bb20})
is the correct answer to (\ref{680a}), which in turn is the wrong
approximation of the original amplitude. The error is hardly surprising,
since the term $\beta _{III}$ has been omitted and the term $\beta _{I}$ has
been approximated in a totally unsystematic way.

To demonstrate that expansion in powers of $\rho $ is a non-starter we go
one step further in the expansion of (\ref{e24b}) retaining terms of order $%
\rho ^{2}$. Denoting this second order approximation to $I$ by $I^{(2)}$ we
get

\begin{equation}
I^{(2)}=-\frac{i\omega }{2}\int_{0}^{\ln 2}d\rho \rho ^{i\omega
+1}e^{i\omega ^{\prime }\rho }  \label{680b}
\end{equation}

\noindent The calculation of (\ref{680b}) proceeds along exactly the same
lines as that of (\ref{bb18}) and yields 
\begin{equation}
I^{(2)}=-\frac{i\omega }{2}\left( \ln 2\right) ^{i\omega +1}\frac{1}{i\omega
+2}M(i\omega +2,i\omega +3,i\omega ^{\prime }\ln 2)  \label{685}
\end{equation}

\noindent In the $\omega ^{\prime }\rightarrow \infty $ limit 
\begin{equation}
M(i\omega +2,i\omega +3,i\left| z\right| )\approx -i\left( 2+i\omega \right) 
\frac{e^{i\left| z\right| }}{\left| z\right| }-\Gamma \left( 3+i\omega
\right) \frac{e^{-\frac{\pi \omega }{2}}}{\left| z\right| ^{2+i\omega }}
\label{bb21}
\end{equation}

\noindent The second term in (\ref{bb21}) ought to be neglected compared to
the second term in (\ref{bb20}) since the former is down by one power of $%
\omega ^{\prime }$. The first term in (\ref{bb21}) is of the same order in $%
\omega ^{\prime }$ as the two terms in (\ref{bb20}) and thus ought to be
retained. The pattern persists to all orders, and is due to the structure $%
M(i\omega +n,i\omega +n+1,i\omega ^{\prime }\ln 2)$ ($n$ integer) of the
confluent in the present problem (the second argument being equal to the
first plus one). Thus an expansion in powers of $\rho $ gives contributions
of the same order of magnitude (i.e. $O\left( 1/\omega ^{\prime }\right) $
and hence is of little use.

In retrospect it may seem quite surprising how one obtains the correct
result starting with the wrong expression (\ref{680a}). Let us return to
that expression which (as detailed above) is the first order approximation
in a short distance expansion from the asymptote. The integral is often
handled as follows (see for example (Birrell and Davies 1982) p. 108. One
rescales the variable and writes the integral in the form 
\begin{equation}
I^{(1)}=\left( \omega ^{\prime }\right) ^{-i\omega }\int_{0}^{\omega
^{\prime }\ln 2}d\rho \rho ^{i\omega }e^{i\rho }  \label{b1}
\end{equation}

\noindent One now simply sets $\omega ^{\prime }\ln 2=\infty $, changes
variable $\rho =i\sigma $ and rotates in the complex plane to get (\ref{b1})
in the form 
\begin{equation}
I^{(1)}\simeq e^{-\frac{\pi \omega }{2}}\left( \omega ^{\prime }\right)
^{-i\omega }\int_{0}^{\infty }d\sigma e^{-\sigma }\sigma ^{i\omega }
\label{e18}
\end{equation}

\noindent Notice that setting $\omega ^{\prime }\ln 2=\infty $ certainly
does {\it not }amount to a systematic expansion in $\left( \omega ^{\prime
}\right) ^{-1}$. The $\sigma $ integration yields $\Gamma (1+i\omega )$ and
one thus obtains the form for the $\beta $ amplitude leading to the black
body spectrum. On the other hand the integral (\ref{680a}) can be performed
exactly (see (\ref{bb18}) above) and the asymptotic estimate for large $%
\omega ^{\prime }$ (recall that we are chasing the ultraviolet divergence)
is given by (\ref{bb20}). The reason for the discrepancy lies in the fact
that one should first evaluate the integral in terms of the confluent and
then take the $\omega ^{\prime }\rightarrow \infty $ limit rather than take
the limit first. This rotation in the complex plane stumbles upon the Stokes
phenomenon for the confluent (different limits for $\left| z\right|
\rightarrow \infty $ depending on $\arg z)$). In other words it appears that
Birrell and Davies have made two self-cancelling mistakes (wrong
approximation to the amplitude and wrong evaluation of the integral).

\end{document}